\documentclass[12pt]{iopart}

\def\bea{\begin{eqnarray}}
\def\eea{\end{eqnarray}}

\def\beq{\begin{equation}}
\def\eeq{\end{equation}}

\newbox\pippobox

\begin{document}

\maketitle

\title{Degeneracy pressure of relic neutrinos and cosmic coincidence problem}

\author{Tae Hoon Lee}
\address{Department of Physics and Institute of Natural Sciences, \\
Soongsil University, Seoul 156-743 Korea} \ead{thlee@ssu.ac.kr}

\date{\today}
\begin{abstract}
 We consider the universe as a huge $\nu_R$-sphere formed
 with degenerate relic neutrinos and
suggest that its constant energy density play a role of an effective
cosmological constant. We construct the sphere as a bubble of true
vacuum in a field theory model with a spontaneously broken $U(1)$
global symmetry, and we interpret the sphere-forming time as the
transition time for recent acceleration of the universe. The
coincidence problem may be regarded as naturally resolved in this
model, because the relic neutrinos can make the $\nu_R$-sphere at
the recent past time during the matter-dominated era.
\end{abstract}



\maketitle

\section{Introduction}

The inflationary big bang cosmology has been developed into a
precision science by recent cosmological observations including
supernova data~\cite{ob} and measurements of cosmic microwave
background radiation~\cite{ob2}. From those it is suggested that our
universe is made up of about 74 percent dark energy as well as about
4 percent ordinary matter and about 22 percent dark matter. The
observations triggered an explosion of recent interests in the
origin of dark energy~\cite{dark}. There are several approaches to
understanding the dark energy such as quintessence~\cite{cald} and
phantom~\cite{phantom}.

One of the simple ways to explain it is through an introduction of
the cosmological constant $\Lambda$, which leads to an exponential
expansion of the scale factor of the universe via $a(t)\sim
e^{\sqrt{\Lambda/3}~t}$. Even though this simple form adequately
describes current acceleration with a small value of $\Lambda$, it
cannot say anything about important issues including the coincidence
problem \cite{coinci}. Why in the recent past the expansion of the universe 
started to accelerate? In this article, we suggest relic
neutrinos, which are the lightest among all known fermionic
particles and thus have become non-relativistic in the last cosmic
time, might be related with this issue. We study a possible role
of degeneracy pressure of the neutrinos in the evolution of the
universe after neutrino decoupling $t_{d}$.


When we consider a sphere of radius $R(t)$ containing all relic
neutrinos (of total number $N(t)=M(t)/m_{\nu}$) at a cosmic time $t$
($\ll t_d$), the balance between degeneracy energy of the
non-relativistic neutrinos, $ {\hbar^2
M^{{5}/{3}}}/({m_{\nu}^{{8}/{3}} R^2})$, and their gravitational
energy, $ {GM^2}/{R} $, gives us such relation of the radius as $ R
(t) = {\hbar^2}/({ G m_{\nu}^{{8}/{3}} M^{{1}/{3}}(t)}) $.
On the other hand, the balance between degeneracy energy of
relativistic neutrinos, $ {\hbar c M^{{4}/{3}}}/({m_{\nu}^{{4}/{3}}
R})$, and the gravitational energy yields the maximum total mass
\cite{ch}
$ M_{c} =(\frac{\hbar c }{ G})^{\frac{3}{2}} m_{\nu}^{-2} $.
 From these the radius of the sphere is bounded below,
\beq R>R_c ~\big(\simeq \frac{\hbar^2}{ c^2 m_{\nu}^2 l_p}\big) \sim
10^{23}\times\big ( \frac{{1 eV}}{  m_{\nu} c^2}\big)^2 ~ cm
\label{a} \eeq with Planck
length $l_p $. 

The matter-dominated era of our universe began at $t_m= 70~{\rm Kyr}
$ and ended at the transition time
 $t_t= 5~ {\rm Gyr}$ \cite{hartnett}, at which the universe started
to accelerate. In the era $t_m<t<t_{t}$, we can neglect the pressure
of the other matter than the degeneracy pressure of relic neutrinos
caused by the Heisenberg uncertainty principle. If we assume that at
a cosmic time $t=t_{\chi}$ a huge bubble of relic neutrinos (which
we call a $\nu_R$-sphere) is formed with a radius $R_{\chi}$
determined from the condition $R_{\chi}(t_{\chi})=R_c$ as a gigantic
Fermion star \cite{TDL}, then the universe's main-frame in the time
can be imagined as made up of these relic neutrinos filled up to the
Fermi energy. Comparing the horizon radius of the present universe
$R_{now}\sim 10^{29}  cm$ (or $R_m\sim 10^{27}cm$ at $t=t_m$) with
the critical value $R_c(\simeq R_{\chi})$ in equation (\ref{a}), we
see that these values are not different from each other if we take
$m_{\nu}\sim 1 - 10^{-3}~ eV$.

From the time $t\simeq t_{\chi}$, the universe has a total energy
density contributed mainly from the $\nu_R$-sphere with a
(effective) cosmological constant $\rho_{\nu}~(\simeq M_c/R_c^3)$ in
addition to cold dark matter $\rho_{CDM}(\propto R^{-3})$, and thus
the time $t_{\chi}$ can be interpreted as the transition time
$t_{t}$.
We think that the coincidence problem might be naturally resolved
because neutrinos become relic in this matter-dominated era and can
make the $\nu_R$-sphere at the recent past time
$t_{\chi}~(<t_{now})$.
 In Section 2 we demonstrate a realization of this idea in a global
$U(1)$ symmetric field theory model with a complex scalar field and
relic neutrinos in curved spacetime. Section 3 contains summary and
discussions.

\section{A Global U(1) Model}

\subsection{General Relativistic Formulation}

The action of a global $U(1)$ symmetric model of complex scalar
$\Phi $ and neutrino fields $\Psi$ minimally coupled to gravity is
given by $S_{m}=\int d^4 x \sqrt{-g}~{\cal L}_{matter}$, where
\begin{eqnarray}
{\cal L}_{matter}&= & -  g^{\mu \nu}\partial_{\mu}\Phi^{\dagger}
\partial_{\nu}\Phi -V (\Phi^{\dagger}\Phi ) \\ \nonumber
&+&\frac{i}{2}(\overline{\Psi} \gamma^a \nabla_a \Psi - \nabla_a
\overline{\Psi} \gamma^a \Psi)-m_{0} \overline{\Psi} \Psi
-q\Phi^{\dagger} \Phi  \overline{\Psi} \Psi \nonumber
\end{eqnarray}
with a bare mass of neutrino $m_{0}$.  The last term describes the
scalar-neutrino interaction 
preserving the $U(1)$ symmetry with a parameter $q$ of dimension,
for example, $1/M_{p}$ (Planck mass $M_p$) in gravity-induced $U(1)$
breaking models \cite{MG, gi}.

From the above action, we obtain following equations for scalar
field and neutrino.
\begin{equation}
\frac{1}{\sqrt{-g}} \partial_{\mu}(\sqrt{-g} g^{\mu \nu}
\partial_{\nu} \Phi ) -\frac{\partial V }{\partial \Phi^{\dagger}}
- q \Phi \overline{\Psi} \Psi =0~ , \label{sc}
\end{equation}
\begin{equation}
i\gamma^a \nabla_a \Psi -m_0 \Psi - q \Phi^{\dagger} \Phi \Psi=0~ ,
\end{equation}
where the $\gamma^a$-matrices satisfy the Clifford algebra in a
locally flat inertial coordinate,
$\{ \gamma^a ,\, \gamma^b  \}=-2\, \eta^{a b}$
with $
\eta^{a b}=Diag(-1,\,1,\,1,\,1) $,
and the covariant derivative
\begin{equation}
\nabla_a ={e^\mu}_a (\partial_\mu +\Gamma_\mu )
\end{equation}
is constructed from the vierbein ${e^\mu}_a$ and spin connection
$\Gamma_\mu$ \cite{ausy}.
The energy-momentum tensor in this model
\begin{eqnarray}
T_{\mu \nu}&=&2\partial_\mu \Phi^{\dagger} \partial_\nu \Phi -g_{\mu
\nu}[\partial^\beta \Phi^{\dagger} \partial_\beta \Phi
+V(\vert\Phi\vert^2) ]\nonumber\\ &-&\frac{i}{4}[ (\overline{\Psi}
\gamma_\mu \nabla_\nu \psi -\nabla_\nu \overline{\Psi} \gamma_\mu
\Psi )+(\mu \leftrightarrow \nu)]
\end{eqnarray}
is obtained by using the standard definition
$T_{\mu \nu}\equiv -\frac{2}{\sqrt{-g}} \frac{\delta S_{m}}{\delta
g^{\mu \nu}} ~ = -\frac{ e_{a \mu }}{ \det \{ e \} }\frac{\delta
S_m}{\delta e_a^{\nu}}$.
In next subsection we construct a concrete model with a specific
potential of the scalar field.

\subsection{A Bubble Universe with $U(1)$ Symmetry Spontaneously Broken }

 Assuming a potential of the scalar field
\begin{equation}
V(\vert\Phi\vert^2)= \lambda \vert\Phi\vert^2(\vert\Phi\vert^2 - v^2
)^2 \label{v}\end{equation}
with dimensionful constants $\lambda$
and $ v $, we consider a spherical bubble of true vacuum $\big< \Phi
\big>=v$ with a spontaneously broken $U(1)$ global symmetry,
immersed in the different (false) vacuum state $\big< \Phi \big>=0$
having the $U(1)$ symmetry.
To construct a field theoretical model in which the ideas discussed
in the (previous) Section 1 can be simply realized, we adopt the
Thomas-Fermi(TF) approximation \cite{TDL, Q, QGM} inside the bubble
including all relic neutrinos of the total number $ N=\int d^3
x\sqrt{-g}\, {\Psi}^{\dagger}\Psi {e^t}_0$ with ${e^t}_0 =
\frac{1}{\delta \alpha}$.
The neutrinos have the effective mass, $m_{eff}=q v^2$ with $\big<
\Phi\big> =v$, and we put $m_0=0$ from now on for the simplicity.
 In the TF approximation there is at each point in space a Fermi sea of
massive neutrinos with the local Fermi momentum $q_F (r)$, and the
number density of neutrinos is given by
\begin{equation}
\big< \Psi^{\dagger} \Psi \big>_{TF} =\frac{2}{(2\pi)^3 }\int d^3
q\, n_q =\frac{{q_F}^3 (r)}{3\pi^2}~ \label{tf}
\end{equation}
with the Fermi distribution $n_q =\theta (q_F -q)$. From Eq.
(\ref{tf}) and Dirac's equation, $i\partial_t \Psi=E \Psi$, we have
the energy density of neutrinos
\begin{eqnarray}
&& \big< \frac{i}{2\delta \alpha}(\overline{\Psi} \gamma^0
\partial_t
 \Psi -\partial_t \overline{\Psi} \gamma^0 \Psi ) {\big>}_{TF} 
=\frac{2}{(2\pi)^3 }\int d^3 q\, n_q\, E (q) \,{e^t}_0 \equiv \rho
 (r) .
\end{eqnarray}
The pressure $p(r)$ of neutrinos is given in equation (13) below.


When we express the spherically symmetric (static) spacetime as
\begin{equation}
ds^2 =-\delta^2 (r) \alpha^2 (r) dt^2 +\frac{ dr^2}{ \alpha^{2} (r)
} +r^2 d\theta^2 +r^2 sin^2 \theta \, d\phi^2 , \label{metric}
\end{equation}
the (background) scalar field equation (\ref{sc}) reads
\begin{eqnarray}
&& \frac{1 }{r^2 \delta}{( r^2 \delta \alpha^2 {\Phi'} )}'
-\lambda\Phi(\vert\Phi\vert^2-v^2)(3\vert\Phi\vert^2-v^2)
- q\Phi\big<\overline{\Psi} \Psi\big>_{TF}  = 0, ~\label{scalar}
\end{eqnarray}
and the Einstein equation, $G_{\mu\nu}=\kappa T_{\mu\nu}$ with
$\kappa=8\pi G$, in this TF approximation, yields
\begin{eqnarray}
&& \frac{{(1-\alpha^2 ) }}{r^2}- \frac{1}{r}(\alpha^2)'
= 
\kappa[  \alpha^2 \vert{\Phi}'\vert^2 +V  + \frac{i}{2\delta
\alpha}\big< \overline{\Psi} \gamma^0
\partial_t \Psi -\partial_t \overline{\Psi} \gamma^0 \Psi\big>_{TF}
],\nonumber
\end{eqnarray}

\begin{eqnarray}
&& -\frac{{(1-\alpha^2 ) }}{r^2} +  \frac{{(\delta^2 \alpha^2 )'
}}{r \delta^2 }
= 
\kappa[\alpha^2 \vert{\Phi}'\vert^2 - V   -\frac{i \alpha}{2} \big<
\overline{\Psi} \gamma^1
\partial_r \Psi -\partial_r \overline{\Psi} \gamma^1 \Psi \big>_{TF}
],\nonumber
\end{eqnarray}

\begin{eqnarray}
&& \frac{1}{2 \delta^2 }{(\delta^2 \alpha^2 )''} -\frac{{(\delta^2
\alpha^2 )'}} {4\delta^2 }\frac{(\delta^2)'}{\delta^2 } +
 \frac{{(\delta^2 \alpha^2 )' }}{r\delta^2 }
- \frac{\alpha^2 }{2r}\frac{(\delta^2)'}{\delta^2 } \nonumber\\ =&&
\kappa[ -\alpha^2 \vert{\Phi}'\vert^2 - V  - \frac{
 i}{2r}\big<\overline{\Psi} \gamma^2 \partial_\theta \Psi
 -\partial_\theta \overline{\Psi} \gamma^2 \Psi\big>_{TF} ]
 ,\nonumber
\end{eqnarray}

\begin{eqnarray}
&& \frac{1}{2 \delta^2 }{(\delta^2 \alpha^2 )''} -\frac{{(\delta^2
\alpha^2 )'}} {4\delta^2 }\frac{(\delta^2)'}{\delta^2 } +
 \frac{{(\delta^2 \alpha^2 )' }}{r\delta^2 }
- \frac{\alpha^2 }{2r}\frac{(\delta^2)'}{\delta^2 } \nonumber\\ = &&
\kappa[  -\alpha^2 \vert{\Phi}'\vert^2- V  - \frac{ i}{2 r}
\big<\frac{1}{sin\theta }(\overline{\Psi} \gamma^3 \partial_\phi
\Psi -\partial_\phi \overline{\Psi} \gamma^3  \Psi  )\big>_{TF} ] ,
\label{eins}
\end{eqnarray}
where ${\Phi}'\equiv \frac{\partial \Phi}{\partial r}, ~
 ...$ ,
have been used for the simplicity.
The results in above equation (\ref{eins}) are consistent with those
in the case of fermion stars \cite{TDL}, when
\begin{eqnarray}
&&\big< \frac{ \alpha}{2 i} (\overline{\Psi} \gamma^1 \partial_r
\Psi -\partial_r \overline{\Psi} \gamma^1  \Psi ) {\big>}_{TF} =
\big<  \frac{ 1}{2 i r}(\overline{\Psi} \gamma^2 \partial_\theta
\Psi -\partial_\theta \overline{\Psi} \gamma^2
\Psi){\big>}_{TF} \nonumber\\
 &&=\big< \frac{ 1}{2 i r\, sin\theta}(\overline{\Psi} \gamma^3
 \partial_\phi \Psi -\partial_\phi \overline{\Psi} \gamma^3 \Psi
 ){\big>}_{TF} = p(r) ,
\end{eqnarray}
%
are assumed in the spherically symmetric spacetime, and from Dirac's
equation and its Hermitian conjugate we have the relation
$\rho(r) - 3~ p(r) = q v^2 \big< \overline{\Psi} \Psi {\big>}_{TF}$.

Now assume that there exist regular solutions of the metric
components, scalar field, and the energy density and pressure of
neutrinos, and we get approximate solutions to above equations
(\ref{scalar})-(\ref{eins}) as
\begin{eqnarray}
\Phi(r)&=&v+  \frac{\rho_0 -3 p_0}{6} r^2+{\cal O}(r^3 )\, , \nonumber \\
\alpha^2 & =&1-\frac{\kappa}{3}\rho_0 r^2 +{\cal O}(r^3 ) ,\\
\delta^2 \alpha^2& =& \{ 1+\kappa(\frac{1}{6}\rho_0 +\frac{1}{2} p_0
)r^2 \} +{\cal O}(r^3 ),\nonumber
\end{eqnarray}
with $\rho_0\equiv\rho(0)$ and $p_0\equiv p(0)$. These solutions are
consistent with the results in fermionic stars with a global
monopole \cite{QGM}.
%
%
On the other hand, to equation (\ref{eins}) with $V\simeq 0$,
$\rho\simeq 0$ and $p\simeq 0$ far away from the horizon of the true
vacuum bubble $r_h\equiv \sqrt{1/\kappa \rho_0} \ll r$, solutions of
the metric components in equation (\ref{metric}) can be taken
asymptotically as
$\alpha^2 \simeq 1 -\frac{2\kappa M}{r}$ and $ \delta^2 \simeq 1 $
with the total mass of all particles inside the bubble universe $M$.
With the metric components we have the asymptotic solution
$\Phi \simeq r^{-2} e^{-\sqrt{\lambda}v^2 r}$. 
%
Under the condition \beq \sqrt{\lambda} v^2 \kappa M =1,
\label{nl}\eeq the solution for $\Phi$ satisfies all the relevant
equations even upto next leading order.

From the effective neutrino mass $m_{eff}=qv^2$ inside the bubble,
we can estimate the true vacuum expectation value as $v^2\sim
m_{eff}M_p$, which gives us $\lambda\simeq M_p^4/(v^4 M^2)\sim
10^{-120}/m_{eff}^2$ when equation (\ref{nl}) is used.
On the other hand, if the potential $V(\vert \Phi\vert^2)$ of the
scalar field is induced by gravity \cite{MG, gi} with the
dimensionful constant $\lambda \simeq v^{n-2}/ M_{p}^n$ for $0\leq
n$, then its magnitude comparable with the above is given as
$\lambda\sim
 10^{-112}\times m_{eff}^2 /(1 ~eV)^4$ for $n=6$, which is the same as
$10^{-120}/m_{eff}^2$ if $m_{eff}\sim 10^{-2}~ eV$.


\section{Summary and Discussions}

Considering a huge $\nu_R$-sphere to be possibly formed at
$t=t_{\chi}$, with degenerate relic neutrinos filled up to the Fermi
energy, we suggest that its constant energy density $\rho_{\nu}$
play a role of an effective cosmological constant. The time
$t_{\chi}$ can be interpreted as the transition time $t_{t}$, and
these lead to the interesting possibility that our model gives one
way of understanding the coincidence problem \cite{coinci}, because
the relic neutrinos can make the $\nu_R$-sphere at the recent past
time $t_{\chi}~(<t_{now})$ during the matter-dominated era. In the
Section 2 we have constructed a spontaneously broken $U(1)$ global
symmetric model, in which the energy density $T_{t t}\equiv
\rho~(\sim \rho_0)$ in the right hand side of the first equation of
(\ref{eins}) can be put as $\rho_{\nu}\sim M_c/R_c^3$ of the
$\nu_R$-sphere considered from Section 1. However the formulation
was constrained in a static case and further study is necessary on
the cosmic time evolution of the scale factor $R(t)$. The predicted
values of the relic neutrino mass, in the range
 $m_{\nu}\sim 1 - 10^{-3}~ eV$
given both in Section 1 and the last line of Section 2, are
anticipated to be tested by future observations.

We would like to make a few comments. In future works, we need a
fully relativistic treatment (and possible numerical methods) in
calculation of the more rigorous value of the critical radius of the
$\nu_R$-sphere in equation (\ref{a}).
For our $U(1)$ model to work with the appropriate (relic) neutrino
mass, the parameter $\lambda$ in the potential of the scalar field
in equation (\ref{v}) should be very small, which might be realized
in some models using an extremely light scalar field \cite{XHe} or
in gravity-induced models {\cite{MG, gi}. We are seeking to find
others models related with ours, in which the global $U(1)$ symmetry
is spontaneously broken with $\big<\Phi\big>\equiv
v=\sqrt{m_{eff}M_p}\sim 100~ T eV$.
 During a very slow phase
transition ($ \lambda \ll 1$) from the false vacuum with
$\big<\Phi\big>=0$ to the true vacuum with $\big<\Phi\big>=v$,
masses of neutrinos look changing from from $0$ to $m_{eff}=q v^2$,
and the $\nu_R$-sphere made of the massive neutrinos could play a
similar role to a effective cosmological constant in our model,
while a growing matter component (neutrinos) in Ref. \cite{wetter},
instead, helps such a field as the cosmon \cite{wetter} responsible
for the dark energy.

\section*{Acknowledgements}

This research was supported by the Basic Science Research Program
through the National Research Foundation of Korea(NRF) funded by
Ministry of Education, Science and Technology(2010-0012692).



\section*{References}

\begin{thebibliography}{0}

\bibitem{ob} Perlmutter S J et al. 1999 {\it Astrophys. J.} {\bf 517} 565;
Riess A G et al. 2007 {\it Astrophys. J.} {\bf 659} 98,
astro-ph/0611572

\bibitem{ob2} Spergel D N et al. (WMAP Collaboration) 2007 {\it Astrophys. J.} Suppl. {\bf 170} 377, astro-ph/0603449

\bibitem{dark} Perlmutter S J, Turner M S and White M J 1999 {\it Phys. Rev. Lett.} {\bf 83} 670, astro-ph/9901052

\bibitem{cald} Caldwell R R, Dave R and Steinhardt P J 1998 {\it Phys. Rev. Lett.} {\bf 80} 1582,
astro-ph/9708069, and references therein

\bibitem{phantom} Caldwell R R 2002 {\it Phys. Lett.} B {\bf 545} 23, astro-ph/9908168; 
Hong S T, Lee J, Lee T H and Oh P 2008 {\it Phys. Rev.} D {\bf 78} 047503, arXiv:0801.3781 [gr-qc] 


\bibitem{coinci} Chimento L P, Jakubi A S, Pavon D and Zimdahl W 2003 {\it Phys. Rev.} D {\bf 67} 083513,
astro-ph/0303145

\bibitem{ch} Chandrasekhar S 1931 {\it Astrophys. J.} {\bf 74} 81

\bibitem{hartnett} Hartnett J G and Oliveira F J 2006 {\it Found. Phys. Lett.} {\bf 19} 519, astro-ph/0603500

\bibitem{TDL} Lee T D and Pang Y 1987 {\it Phys. Rev.} D {\bf 35} 3678

\bibitem{MG} Barbieri R, Ellis J R, and Gaillard M K 1980 {\it Phys. Lett.} B
{\bf 90} 249

\bibitem{gi}
Akhmedov E, Berezhiani Z, and Senjanovic G 1992 {\it Phys. Rev.
Lett.} {\bf 69} 3013 

\bibitem{ausy}
Lee T H and McKellar B H J 2003 {\it Phys. Rev.} D {\bf 67} 103007

\bibitem{Q}
Bachall S, Lynn B W and Selipsky S B 1989 {\it Nucl. Phys.} B {\bf
325} 606

\bibitem{QGM}
Li X Z and X. H. Zhai X H 1995 {\it Phys. Lett.} B {\bf 364} 212

\bibitem{XHe}
He X G, Mckellar B H J and Stephenson Jr. G J 2004 {\it Phys. Lett.}
B {\bf 444} 75, Erratum-ibid. B {\bf 581} 270 (2004), hep-ph/9807338

\bibitem{wetter}
Amendola L, Baldi M, and Wetterich C 2008 {\it Phys. Rev.} D {\bf
78} 023015, arXiv:0706.3064 [astro-ph]


\end{thebibliography}
\end{document}